\documentclass[twocolumn,letter]{jpsj3}
\usepackage{txfonts}
\usepackage{graphicx}
\usepackage{dcolumn}
\usepackage{bm}
\usepackage{color}

\newcommand{\beq}{\begin{eqnarray}}
\newcommand{\eeq}{\end{eqnarray}}

\begin{document}

\title{
Fulde-Ferrell-Larkin-Ovchinnikov States in Two-Band Superconductors
}

\author{Takeshi Mizushima$^{1,2}$\thanks{E-mail: mizushima@mp.okayama-u.ac.jp}, 
Masahiro Takahashi$^3$, and Kazushige Machida$^1$}
\inst{$^1$Department of Physics, Okayama University, Okayama 700-8530, Japan \\
$^2$Department of Physics and Astronomy, Northwestern University, Evanston, Illinois 60208, USA \\
$^3$Department of Physics, Gakushuin University, Tokyo 171-8588, Japan} %\\

%\author{Takeshi Mizushima}
%\email{mizushima@mp.okayama-u.ac.jp}
%\affiliation{Department of Physics, Okayama University, Okayama 700-8530, Japan} 
%\affiliation{Department of Physics and Astronomy, Northwestern University, Evanston, Illinois 60208, USA}
%\author{Masahiro Takahashi}
%\affiliation{Department of Physics, Gakushuin University, Tokyo 171-8588, Japan}
%\author{Kazushige Machida}
%\affiliation{Department of Physics, Okayama University,
%Okayama 700-8530, Japan}
\date{\today}

%\begin{abstract}
\abst{
We examine the possible phase diagram in an $H$-$T$ plane for Fulde-Ferrell-Larkin-Ovchinnikov (FFLO) states in a two-band Pauli-limiting superconductor. We here demonstrate that, as a result of the competition of two different modulation length scales, the FFLO phase is divided into two phases by the first-order transition: the $Q_1$- and $Q_2$-FFLO phases at the higher and lower fields. The $Q_2$-FFLO phase is further divided by successive first order transitions into an infinite family of FFLO subphases with rational modulation vectors, forming a {\it devil's staircase structure} for the field dependences of the modulation vector and paramagnetic moment. The critical magnetic field above which the FFLO is stabilized is lower than that in a single-band superconductor. However, the tricritical Lifshitz point $L$ at $T_{\rm L}$ is invariant under two-band parameter changes. 
}
%\end{abstract}
%\kword{multiband superconductivity, Fulde-Ferrell-Larkin-Ovchinnikov states, successive first-order phase transitions, devil's staircase structure}

\setlength{\textheight}{670pt}

%\pacs{74.20.-z, 74.25.Dw, 74.81.-g } %74.20.-z, 74.81.-g, 74.25.Dw}

%74.20.Rp Pairing symmetries (other than s-wave)
%74.20.-z Theories and models of superconducting state
%74.25.Dw Superconductivity phase diagrams
%74.81.-g Inhomogeneous superconductors and superconducting systems, including electronic inhomogeneities

%74.70.Ad Metals; alloys and binary compounds (including A15, MgB2, etc.)

\maketitle

%---------- Introduction
{\it Introduction.---}
Owing to the fundamental significance of the coexistence of superconductivity and magnetism, Fulde-Ferrell-Larkin-Ovchinnikov (FFLO) states~\cite{ff,lo} have attracted much attention in the fields of condensed matter~\cite{review_sc}, cold atoms~\cite{review_atom,TM2005,mit,rice,cai}, and neutron stars~\cite{review_nuetron}. Still FFLO remains elusive in spite of extensive experimental and theoretical investigations in a wide range of fields. The emergence of the FFLO state via the second-order transition is accompanied by the Jackiw-Rebbi soliton~\cite{jackiw} by which the Pauli paramagnetic moment is neatly accommodated~\cite{machida1984}. The soliton provides a generic key concept for the common understanding of the essential physics of the FFLO phase in a single-band superconductor, incommensurate structures~\cite{ssh,takayama}, and fermionic excitations bound at topological defects of a superconductor~\cite{TM2005v2}. 

%Mobile electrons do not neccessarily live in a single band, but may be in multiple sheets of Fermi surfaces. 

Multiband effects with multiple sheets of Fermi surfaces are observed in a variety of superconductors, such as MgB$_2$, iron-based superconductors, and Sr$_2$RuO$_4$ to mention a few~\cite{komendova,review214}. It has been proposed that multi-band effects are accompanied by exotic superconductivity.~\cite{babaev,babaev2,hirano,kogan2011} It is now recognized that the multiband superconductor is the rule rather than the exception. Nevertheless, in contrast to those in a single-band superconductor~\cite{review_sc}, the FFLO phases in multiband systems have not been clarified so far, except in few studies~\cite{gurevich,ptok,ptok2}.

%Needless to say, background electrons do not necessarily live in a single band, but may be in multiple sheets of Fermi surfaces. In actual, multiband effects have been observed in various superconducting materials, such as MgB$_2$, Iron-based superconductors, and xxx~\cite{komendova}. Nevertheless, in contrast to extensive study in a single-band superconductor, FFLO phases in multiband systems have not been clarified so far. 

\begin{figure}[b]
\begin{center}
\includegraphics[width=80mm]{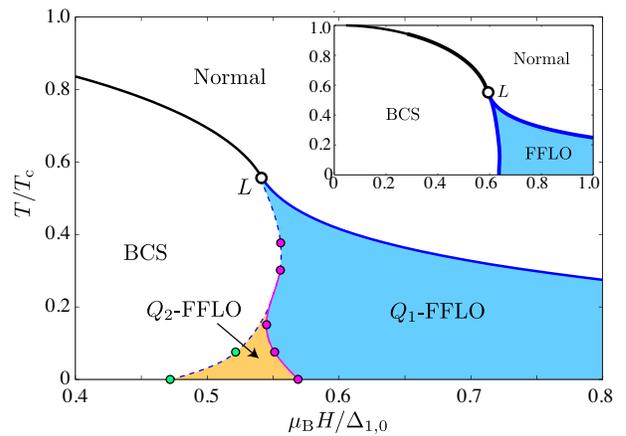}
\end{center}
\caption{(color online) Phase diagram in an $H$-$T$ plane. The thick (thin) lines indicate the second (first) -order transition lines. The $Q_2$-FFLO phase is subdivided into a family of FFLO subphases. The open circle is the Lifshitz point $L$ at $(T_{\rm L}, H_{\rm L})$. The order of the dashed line is undetermined. The inset describes the phase diagram in a single-band system~\cite{machida1984}.}
\label{fig:phase}
\end{figure}

Results of several recent experiments have collectively urged us to investigate the FFLO phases in multibands: The observations of strong Pauli effects in iron pnictides~\cite{cho,terashima,burger,zocco,kittaka} and of a strange phase boundary line, $\frac{dT_{\rm LO}}{dH} \!>\! 0$ ($T_{\rm LO}$ is the BCS-FFLO transition line), in another multiband heavy-fermion superconductor, CeCoIn$_5$, for $H \!\parallel\! ab$~\cite{kenzelmann}, which is in contrast with a conventional phase diagram with $\frac{dT_{\rm LO}}{dH}\!<\! 0$ (see the inset of Fig.~\ref{fig:phase} and Refs.~\cite{review_sc} and \cite{suzuki}). 

In a single-band superconductor, the self-organized periodic structure of the FFLO state is a direct consequence of the synergistic effect between spin paramagnetism and superconductivity with the spontaneous breaking of the translational symmetry. The spatial modulation is characterized by the single length scale $Q^{-1}$, proportional to the Fermi velocity $Q^{-1}\!\propto\! v_{\rm F}$~\cite{machida1984}. Multiband superconductivity is characterized by multicomponent pair potentials. In the absence of interband coupling, the $\gamma$ band ($\!=\! 1, 2$) independently has its own favored FFLO modulation $Q^{-1}_{\gamma}$. When the interband coupling becomes finite, however, two pair potentials are no longer independent and the coupling gives rise to the competition of multiple length scales self-consistently determined, $Q^{-1}_{\gamma}$. %In general, the ordered phase under multiple competing length scales yields commensurate-incommensurate structures~\cite{bak,chaikin} and type-1.5 superconductivity~\cite{babaev} between type-I and II as a compromise. 

In this Letter, we try to establish the essential features of FFLO characteristic of multiband superconductors, which are absent in a single-band case~\cite{machida1984}, and examine the possible phase diagram for FFLO phases. The resulting phase diagram is summarized in Fig.~\ref{fig:phase}, where our main outcome is threefold: (i) The FFLO phase is divided into two main phases by the first-order transition, each having a different modulation periodicity. The $Q_1$-FFLO phase stabilized in the higher field is understandable with the single modulation vector $Q_1$ of the major band ($\gamma = 1$). The $Q_2$-FFLO phase in Fig.~\ref{fig:phase} is a result of the competing effect of two different length scales $Q^{-1}_1\!\neq\!Q^{-1}_2$ and is unique to two-band superconductors. (ii) The $Q_2$-FFLO phase is further subdivided by first-order transitions into a family of FFLO subphases with the sequence of rational modulation vectors $Q_2/(2n+1)$ ($n\!\in\! \mathbb{Z}$). This successive first-order phase transition exhibits a {\it devil's staircase structure} for the modulation period and magnetization. (iii) The onset field of the FFLO phase is lower than the Lifshitz point $H_{\rm L}$ owing to the interband effect and the phase boundary indicates a positive slope $\frac{dT_{\rm LO}}{dH} \!>\! 0$.

%-------------------- 
{\it Formulation.---}
We consider spin-$\frac{1}{2}$ fermions ($\sigma \!=\! \uparrow, \downarrow$) in two bands $\gamma \!=\! 1, 2$ under a magnetic field $H$, interacting through an attractive $s$-wave interaction $g_{\gamma\gamma^{\prime}} \!=\! g _{\gamma^{\prime}\gamma} \!<\! 0$. We deal with a quasi-one-dimensional (Q1D) system along the FFLO modulation vector ($\hat{\bm z}$-axis). This is a minimal extension of a single-band theory~\cite{machida1984,yoshii}. The quasiparticles with the wave function ${\bm \varphi}_{\nu,\gamma} \!=\! [u_{\nu,\gamma},v_{\nu,\gamma}]^{\rm T}$ and energy $E_{\nu}$ in the $\gamma$-band are obtained by solving the Bogoliubov-de Gennes (BdG) equation~\cite{suhl}
\beq
\left[
\begin{array}{cc}
\xi _{\gamma} (z) & \Delta _{\gamma}(z) \\ \Delta^{\ast}_{\gamma}(z) 
& - \xi^{\ast}_{\gamma} (z)
\end{array}
\right]{\bm \varphi}_{\nu,\gamma}(z) = E_{\nu,\gamma}
{\bm \varphi}_{\nu,\gamma}(z),
\label{eq:bdg}
\eeq
where ${\bm \varphi}_{\nu,\gamma}$ must fulfill $\int dz{\bm \varphi}^{\dag}_{\nu,\gamma}{\bm \varphi}_{\nu,\gamma} \!=\! 1$. In this paper, we set $\hbar\!=\! k_{\rm B} \!=\! 1$. The single-particle Hamiltonian density is 
$
\xi_{\gamma}(z) \!=\! 
\frac{1}{2M_{\gamma}}(-i\frac{d}{dz}+A)^2 - \mu _{\gamma} - \mu _{\rm B}H$ 
with the mass $M_{\gamma}$. The chemical potentials $\mu _{\gamma}$ are parameterized as $\mu _1 \!=\! E_{\rm F} + \mu_{12}/2$ and $\mu _2 \!=\! E_{\rm F} - \mu _{12}/2$. Since we are interested in the strong Pauli limit, the vector potential $A$ is supposed to be spatially uniform, neglecting the orbital effect~\cite{suzuki}. Note that the orbital effect can be suppressed in low-dimensional superconductors~\cite{yonezawa,bergk}.

The BdG equation Eq.~(\ref{eq:bdg}) is self-consistently coupled with the gap equation in the $\gamma$-band given by
\beq
\Delta_{\gamma}(z) = \sum _{\gamma^{\prime}} g_{\gamma\gamma^{\prime}}\Phi _{\gamma^{\prime}}(z),
\label{eq:gap}
\eeq
where $\Phi _{\gamma}(z) = \sum _{\nu} u_{\nu,\gamma}(z)v^{\ast}_{\nu,\gamma}(z)f(E_{\nu,\gamma})
$ describes the Cooper pair amplitude in the $\gamma$-band with the distribution function at the temperature $T$, $f(E) \!=\! 1/(e^{E/T}+1)$. % and the sum in $\Phi _{\gamma}$ is taken over all quasiparticle states with positive and negative energies. 

In Eq.~(\ref{eq:gap}), $g_{12} \!=\! g_{21}$ denotes the amplitude of the interband pair tunneling. Although we examined several sets of parameters $(g_{12}/g_{11},g_{22}/g_{11},{\mu}_2/\mu _1)$~\cite{takahashi}, we here concentrate on the set $( 0.1, 0.6, 0.5)$, where the density of states (DOS) at the Fermi surface of normal electrons, $\mathcal{N}_{\gamma}$, is dominated by the minor band $\mathcal{N}_2/\mathcal{N}_1 \!=\! 2$. Thus, the Fermi velocity ratio is $v_{{\rm F},2}/v_{{\rm F},1} \!=\! 1/\sqrt{2}$ in our 1D parabolic dispersion. We here assume $M_1 = M_2$ because the deviation merely alters the ratios of $v_{{\rm F},2}/v_{{\rm F},1}$ and $\mathcal{N}_2/\mathcal{N}_1$. This results in $\Delta _{2,0}/\Delta _{1,0} \!\approx\! 0.5$ at $T\!=\! H \!=\! 0$, where $\Delta _{\gamma,0} \!\equiv\! \Delta _{\gamma}(T\!=\!0)$. Namely, the band $\gamma \!=\! 1$ ($=\! 2$) is major (minor) in its gap.

We here consider a 1D modulation with the period $L$, 
\beq
\Delta _{\gamma}(z + L/2) = e^{i\chi}\Delta _{\gamma}(z),
\label{eq:bc}
\eeq
where $\chi \!=\! \pi$ ($2\pi$) corresponds to FFLO (BCS) states. This imposes the periodic boundary condition on quasiparticle wave functions, 
%\beq
${\bm \varphi}_{\nu,\gamma}( z + pL/2) 
\!=\! e^{ikR}e^{i\chi \sigma _z/2}{\bm \varphi}_{\nu,\gamma}(z)
$, 
%\eeq
where $k \!=\! \frac{2\pi q}{LN_{L}}$ is the Bloch vector and $R$ denotes the Bravais lattice vector that satisfies $kR \!=\! \pi pq/N_L$ with $p, q \!\in\! \mathbb{Z}$. Hence, we self-consistently solve Eq.~(\ref{eq:bdg}) coupled with the gap equation (\ref{eq:gap}) in the interval $z \!\in \! [0,L/2]$. The BdG equation Eq.~(\ref{eq:bdg}) is numerically diagonalized with the finite element method implemented with the discrete variable representation.~\cite{takahashi} In this work, we deal with $L/\xi \!<\! 40$, where $\xi \!=\! v_{{\rm F},1}/\Delta _{1,0}$ is the coherence length.

%-------------------- 
{\it Sequence of FFLO states.---}
Within the condition in Eq.~(\ref{eq:bc}), there exists a family of FFLO states as a consequence of the interplay between two bands. To clarify this, we start with the Fourier expansion in FFLO states, $\Delta _{\gamma} (z) \!=\! \sum _{m_{\gamma}\in \mathbb{Z}} e^{iq_{\gamma}(2m_{\gamma}+1)z} \Delta^{(m_{\gamma})}_{\gamma}$. Note that the symmetry requires $Q(2m-1)\!\equiv\!q_1 (2m_1-1) \!=\! q_2(2m_2-1)$. Then, $\Delta _{1,2}(z)$ is expanded with $Q \!=\! \frac{2\pi}{L}$ as
\beq
\Delta _{\gamma} (z) 
= \sum _{m \in\mathbb{Z}} e^{iQ(2m-1)z}\Delta^{(m)}_{\gamma}.
\label{eq:four}
\eeq
In a single-band superconductor with $\Delta _2 \!=\! 0$, an isolated kink state characterized by $\Delta^{(m)}_1 \!=\! \frac{\Delta _{1,0}}{2|m|-1}$ is stabilized at the critical field $\mu _{\rm B}H \!=\! \frac{2}{\pi}\Delta _{1,0}$.~\cite{machida1984} %This is the one-soliton formation energy~\cite{machida1984}. 
The higher Fourier components with $|m| \!\ge\! 2$ disappear as $H$ increases and the spatial modulation results in the sinusoidal form $\Delta _1 (z) \!\propto\! \sin(Qz)$. The field dependence of $Q$ follows the relation $Q \!\sim\! 2\mu _{\rm B}H/v_{\rm F}$ in the high-field limit~\cite{machida1984}. In the case of two-band superconductors, the modulation vector $Q_{\gamma}$ of $\Delta _{\gamma}(z)$ is determined as a result of the competition between two bands, where the $\gamma \!=\! 1$ ($\gamma \!=\!2$) band favors the modulation vector $Q_1 \!\propto\! v^{-1}_{{\rm F},1} \!\propto\! \mu^{-1}_1$ ($Q_2 \!\propto\! v^{-1}_{{\rm F},2} \!\propto\! \mu^{-1}_2$) and $Q_1 \!<\! Q_2$ in our system.

\begin{figure}[t!]
\begin{center}
\includegraphics[width=80mm]{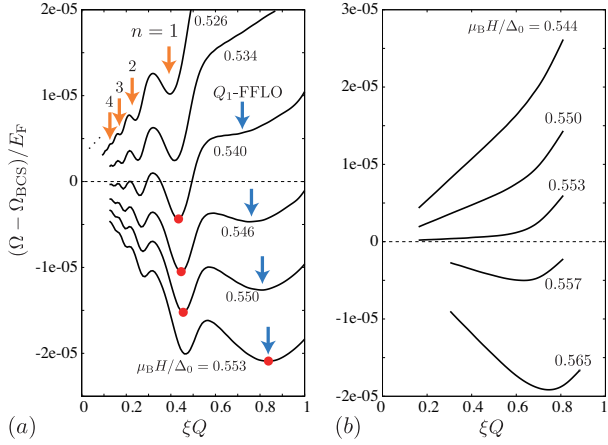}
\end{center}
\caption{(color online) Thermodynamic potential $\Omega$ with respect to the FFLO period $Q\!\equiv\! 2\pi/L$ at $T/T_{\rm c}\!=\! 0.075$ (a) and $0.30$ (b) for various magnetic fields $H$.}
\label{fig:omega}
\end{figure}

Figure~\ref{fig:omega} shows the thermodynamic potential 
$\Omega \!=\! -
\sum _{\gamma,\gamma^{\prime}}g_{\gamma\gamma^{\prime}}\langle \Phi^{\ast}_{\gamma}(z)\Phi _{\gamma^{\prime}}(z)\rangle
+ \sum _{\nu,\gamma} \{ E_{\nu,\gamma} 
\langle | u_{\nu,\gamma}|^2\rangle 
- T\ln(1 + e^{- E_{\nu,\gamma}/T} ) \}$ for a fixed FFLO period $Q^{-1}\equiv L/2\pi$, 
where $\langle \cdots \rangle$ denotes the spatial average over the system. We evaluate $\Omega(Q)$ with self-consistent solutions of Eqs.~(\ref{eq:bdg}) and (\ref{eq:gap}). It is seen from Fig.~\ref{fig:omega}(a) that $\Omega (Q)$ has several local minima, $\frac{\partial \Omega}{\partial Q} \!=\! 0$ and $\frac{\partial^2\Omega}{\partial Q^2} \!\ge\! 0$, in the lower-temperature regime. The local minimum with the largest $Q$ ($\xi Q \!\sim\! 0.8$) corresponds to $Q_1 \xi$, which is favored by the major band, and the other minima with small $Q$'s follow $Q_2/(2n+1)$ with $n\!\ge\! 1$. This is in contrast to the higher-temperature regime shown in Fig.~\ref{fig:omega}(b), where only a single minimum exists, which monotonically shifts towards the shorter $L$ as $H$ increases. The FFLO period $Q$ tends to $Q_1 \!\sim\! \mu _{\rm B}H/v_{{\rm F},1}$ at high fields, which is understandable from the single-band picture.

\begin{figure}[t!]
\begin{center}
\includegraphics[width=80mm]{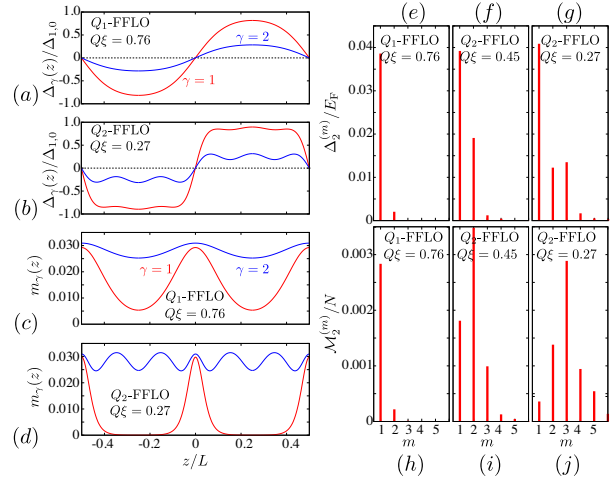}
\end{center}
\caption{(color online) Spatial profiles of $\Delta _{\gamma}(z)$ and $m_{\gamma}(z)$ in $Q_1$- and $Q_2$-FFLO states at $T/T_{\rm c0}\!=\! 0.075$ and $\mu _{\rm B}H/\Delta _{1,0} \!=\! 0.546$: $Q\xi \!=\! 0.76$ (a, c) and $0.27$ (b, d). Histogram of corresponding Fourier components $\Delta^{(m)}_{2}$ (e-g) and $\mathcal{M}^{(m)}_2$ (h-j) in the $\gamma \!=\! 2$ band.}
\label{fig:gap}
\end{figure}

To clarify the distinction between the $Q_1$- and $Q_2$-FFLO states, we show $\Delta _{\gamma}(z)$ at $\mu _{\rm B}H/\Delta _{1,0} \!=\! 0.546$ and $T\!=\! 0.075T_{\rm c}$ in Figs.~\ref{fig:gap}(a) and \ref{fig:gap}(b), which correspond to the local minima with the largest and second largest $Q$ values, respectively. Figures \ref{fig:gap}(e)-\ref{fig:gap}(g) show the Fourier components $\Delta^{(m)}_{2}$ in Eq.~(\ref{eq:four}). It is seen from Figs.~\ref{fig:gap}(a) and \ref{fig:gap}(e) that the local minimum state with $Q\xi \!=\! 0.76$ is characterized by a single peak at $m\!=\! 1$, corresponding to $Q \!\equiv\! \frac{2\pi}{L} \!\sim\! Q_1$. In this sense, we refer to this phase as the $Q_1$-FFLO phase. In the $Q_1$-FFLO phase, the Pauli paramagnetic moment $\mathcal{M}_{\gamma} \!=\! \langle m_{\gamma}\rangle \!=\! \sum _{\nu}[\langle |u_{\nu,\gamma}|^2\rangle f(E_{\nu,\gamma})-\langle |v_{\nu,\gamma}|^2\rangle f(-E_{\nu,\gamma})]$ in the $\gamma$-band accumulates in the FFLO node at which the midgap bound states with spin $\uparrow$ ($\downarrow$) are formed inside (outside) of the Fermi surface as the Jackiw-Rebbi soliton. Hence, the spatial modulation of $\mathcal{M}_{\gamma = 2}$ is characterized by a single $2Q$, as shown in Figs.~\ref{fig:gap}(c) and \ref{fig:gap}(h), where the spatial uniform contribution of $\mathcal{M}_{\gamma}$ with $n \!=\! 0$ is omitted. Here, $\mathcal{M}_{\gamma}(z)$ is expanded with $Q \!=\! \frac{2\pi}{L}$ as $\mathcal{M}_{\gamma}(z) \!=\! \sum _{m \in \mathbb{Z}}e^{i2mQz}\mathcal{M}^{(m)}_{\gamma}$.

In the low-$T$ regime, the contributions of the minor band become competitive, giving rise to the appearance of several local minima in addition to the $Q_1$-FFLO state, as shown in Fig.~\ref{fig:omega}(a). This competing effect is also reflected in the phase diagram shown in Fig.~\ref{fig:phase}, where the critical field above which the $Q_2$-FFLO phase appears is much lower than $h_{\rm cri} \!=\! \frac{2}{\pi}\Delta$ in a single-band system~\cite{machida1984}. To understand the structure of the $Q_2$-FFLO phase, in Figs.~\ref{fig:gap}(b) and \ref{fig:gap}(g), we display the spatial profile of $\Delta _{1,2}(z)$ with $Q\xi \!=\! 0.27$ at $\mu _{\rm B}H/\Delta _0 \!=\! 0.546$, corresponding to the local minimum labeled as $n\!=\! 2$ in Fig.~\ref{fig:omega}(a). In the regime around the critical field in two-band systems, $Q^{-1}_2$ is comparable to the coherence length $Q^{-1}_2\!\sim \! \xi$, whereas $Q^{-1}_1 \!\gg\! \xi$, since $v_{{\rm F},1} \!>\!v_{{\rm F},2}$. The FFLO phase with the single modulation vector $Q_2$ is not favorable because of the loss of condensation energy, and the stability of the $Q_1$-FFLO phase requires a higher magnetic field. As shown in Eq.~(\ref{eq:four}), however, it is possible to realize a family of $Q_2$'s, such as $Q_2/3, Q_2/5, \cdots$ in two-band systems. It is clearly seen from Figs.~\ref{fig:gap}(b) and \ref{fig:gap}(g) that $\Delta _{1,2}(z)$ with $Q\xi \!=\! 0.27$ is composed of multiple modulation vectors, $3Q \!\sim\! 0.91 \xi^{-1}$ ($m\!=\! 2$) and $5Q\!\sim\! 1.4 \xi^{-1}$ ($m\!=\! 3$) in addition to $Q \!\sim\! 0.27 \xi^{-1}$ ($m\!=\! 1$). Although the modulation vector favored in the minor band is estimated as $Q_2 \!\approx\! 1.4\xi^{-1}$, the optimal wave number $Q\xi \!=\! 0.27$ that determines the overall FFLO period corresponds to $Q\!\sim\!Q_2/5$, and the induced components $3Q$ and $5Q$ are found to be $3Q_2/5$ and $Q_2$, respectively. 

In the series of local minima labeled as $n\!=\! 1, 2, \cdots$ in Fig.~\ref{fig:omega}(a), $\Delta _{\gamma} (z)$ are composed of the overall modulation vector $Q\!\approx\!Q_2/(2n+1)$ and the induced components $(2m-1)Q_2/(2n+1)$ with $m \!=\! 1, 2, \cdots$. This family of $Q \!\approx\! Q_2/(2n+1)$ is referred to as the $Q_2$-FFLO phase, which can be stabilized in the lower-$H$ and lower-$T$ regimes in Fig.~\ref{fig:phase}. The multiple-$Q$ modulated structure in the $Q_2$-FFLO phase is clearly reflected in the spatial profile of the Pauli paramagnetic moment displayed in Fig.~\ref{fig:gap}(d), and the Fourier components in Figs.~\ref{fig:gap}(i) and \ref{fig:gap}(j) have sharp peaks at $2Q_2$. It is also seen from Fig.~\ref{fig:omega}(a) that the family of the $Q_2$-FFLO phase undergoes the first-order transition to the $Q_1$-FFLO phase as $H$ increases.

\begin{figure}[t]
\begin{center}
\includegraphics[width=80mm]{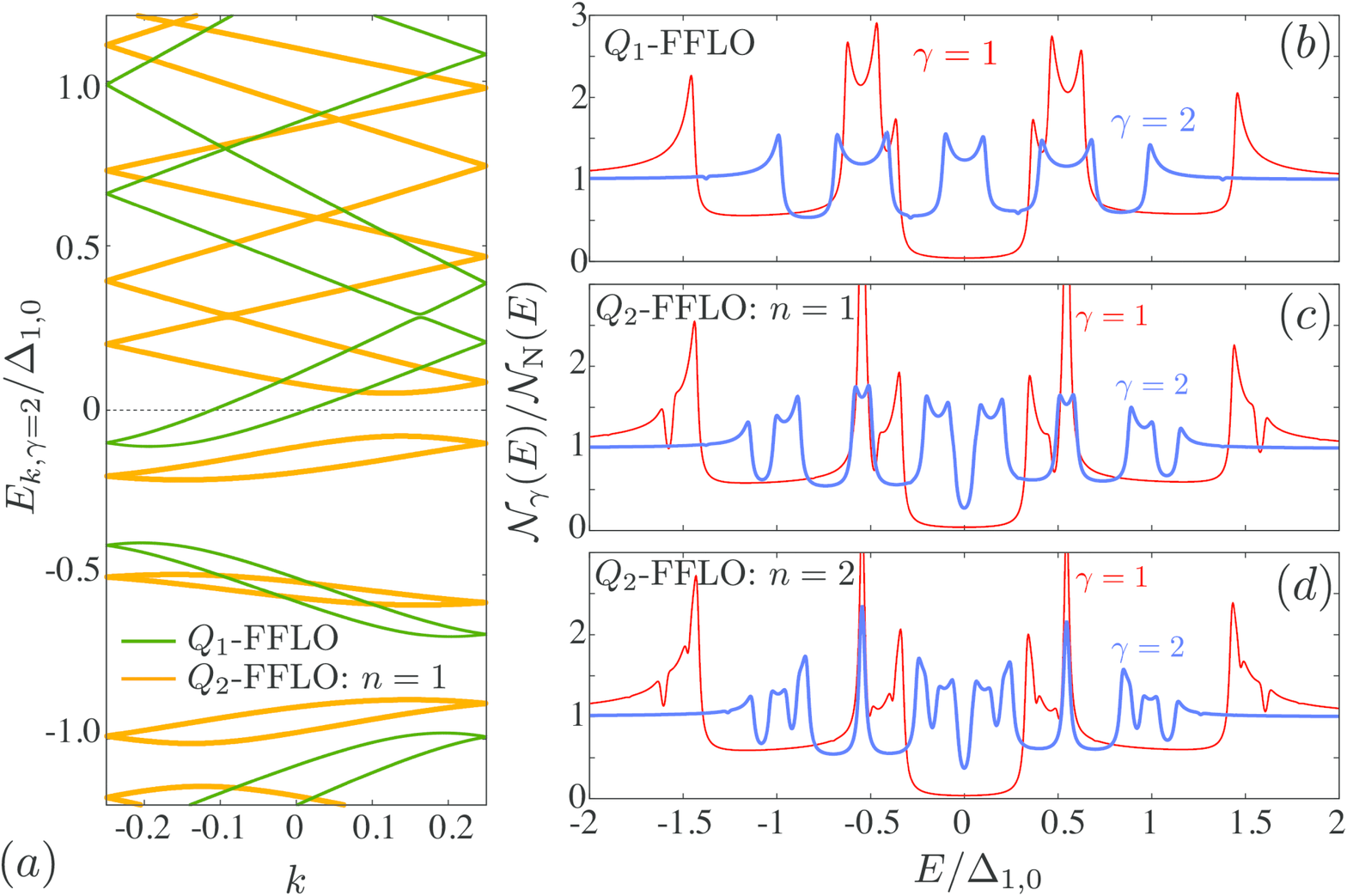}
\end{center}
\caption{(color online) (a) Dispersion in the reduced zone for the $\gamma \!=\! 2$ band of the $Q_1$-FFLO state with $Q\xi \!=\! 0.76$ and $n\!=\! 1$ $Q_2$-FFLO state with $Q\xi \!=\! 0.45$. DOS of $Q_1$-FFLO state with $Q\xi \!=\! 0.76$ (b) and $Q_2$-FFLO state with $Q\xi \!=\! 0.45$ (c) and $Q\xi \!=\! 0.27$ (d). The other parameters are the same as those in Fig.~\ref{fig:gap}}
\label{fig:dos}
\end{figure}

%-------------------- 
{\it Devil's staircase structure.---}
The appearance of the $Q_2$-FFLO phase is a consequence of the competition of the two length scales $Q^{-1}_1$ and $Q^{-1}_2$, which is unique to multiband superconductors. We here discuss the thermodynamic stability of the $Q_1$ and $Q_2$ phases with respect to $H$. Figure~\ref{fig:dos}(a) shows the quasiparticle dispersion in the minor band ($\gamma \!=\! 2$) of the $Q_1$-FFLO state with $Q\xi \!=\! 0.76$ and the $n\!=\! 1$ $Q_2$-FFLO state with $Q\xi \!=\! 0.45$ at $T\!=\! 0.075T_{\rm c}$ and $\mu _{\rm B}H\!=\! 0.546\Delta _{1,0}$. The band structure at approximately $E_{k,\gamma} \!=\! -\mu _{\rm B}H$ is interpreted as the lattice of the Jackiw-Rebbi solitons bound at the FFLO nodes, responsible for the paramagnetic moment. 
 
%Fugure~\ref{fig:dos}(a) shows the dispersion relation, where the band structure around $E_{k,\gamma} \!=\! -\mu _{\rm B}H$ is interpreted as the lattice of the Jackiw-Rebbi solitons, responsible for the paramagnetic moment. The wave function is bound at each FFLO node within the coherence length $\xi$. In the $Q_1$-FFLO phase, since the distance $\frac{\pi}{Q}$ between FFLO nodes are comparable with $\xi$, the band structure of the soliton lattice becomes dispersive. In $Q_2$-FFLO phases, however, the FFLO period $(2n+1)/Q_2$ can be much longer than the length scale of the soliton $\sim\xi$. Hence, the dispersion gets flat as $n$ increases. 

The most distinct structure between the $Q_1$- and $Q_2$-FFLO phases is seen around $E_{k,\gamma}\!=\! 0$. In the $Q_1$-FFLO phase, the dispersion crosses $E_{k,\gamma}\!=\! 0$, giving rise to a large amount of the zero-energy DOS and paramagnetic moment as shown in Figs.~\ref{fig:gap}(b) and \ref{fig:dos}(b). The DOS is defined as 
$\mathcal{N}_{\gamma}(E) \!=\! \sum _{\nu}[\langle |u_{\nu,\gamma}|^2\rangle \delta(E-E_{\nu,\gamma})+\langle |v_{\nu,\gamma}|^2\rangle \delta(E+E_{\nu,\gamma})]$. In the case of the $Q_2$-FFLO phase with $n \!=\! 1$, $\Delta _{2}(z)$ is mostly composed of two different Fourier components, $Q_2/3$ and $Q_2$, as shown in Fig.~\ref{fig:gap}(f), while $\Delta _2(z)$ in the $Q_1$-FFLO phase is described by a single $Q$. Hence, the original band in the $Q_2$-FFLO state with $Q \!=\! Q_2/3$ is folded back into a small reduced Brillouin zone, reflecting the mixed component with the larger modulation vector $3Q \!=\! Q_2$. Then, the band gap opens at approximately zero energy, which reduces $\mathcal{N}_2(E\!=\! 0)$ but gains condensation energy. Hence, the energetics of the FFLO phases is simply understandable as the competition between the $Q_1$-FFLO phase with zero-energy DOS and the $Q_2$-FFLO phase with condensation energy.

\begin{figure}[t!]
\begin{center}
\includegraphics[width=80mm]{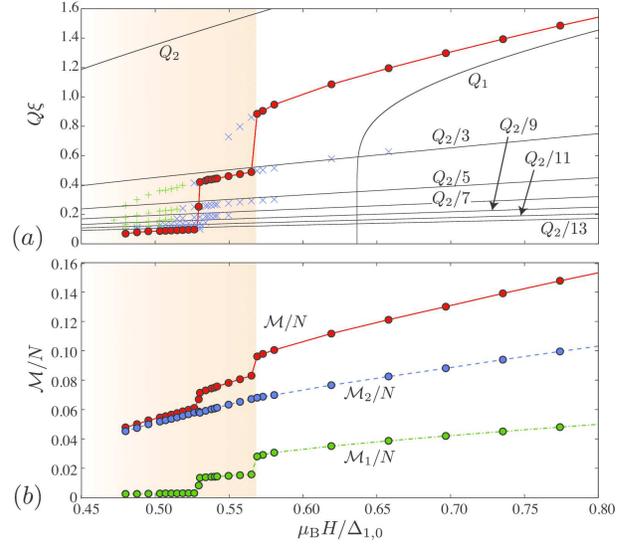}
\end{center}
\caption{(color online) Field dependence of the FFLO modulation $Q$ (a) and the magnetization $\mathcal{M}/N$ (b) at $T\!=\! 0$. The symbols $\times$ and $+$ indicate the metastable states in $\Omega(Q)$, where $\Omega$ in $+$ is higher than that in the BCS phase. The shaded area denotes the $Q_2$-FFLO phase, where the transition between BCS and $Q_2$-FFLO phases takes place at approximately $\mu _{\rm B} H/\Delta _0 \!\approx\! 0.45$.}
\label{fig:phys}
\end{figure}

Figure~\ref{fig:phys}(a) shows a summary of the field dependence of the modulation vector $Q\!\equiv\! 2\pi/L$ at $T\!=\! 0$, where the symbols ``$\times$'' and ``$+$'' denote the metastable solutions. The ground state denoted by filled circles is determined by minimizing $\Omega (L)$. The field dependence of $Q_{1,2}$ in a single-band superconductor is plotted as a reference, where $Q_1$ and $Q_2$ are estimated using ($v_{{\rm F},1}$, $\Delta _{1,0}$) and ($v_{{\rm F},2}$, $\Delta _{2,0}$), respectively~\cite{machida1984}. It is shown in Fig.~\ref{fig:phys}(a) that the $Q(H)$ curve is understandable with two competing modulation vectors, $Q_1$ and $Q_2$. The branch with the largest $Q$, called the $Q_1$-FFLO phase, follows $Q_1(H)$, which is dominated by the $\gamma \!=\! 1$ band with $v_{{\rm F},1}$ and $\Delta _{1,0}$, whereas the branches with the smaller $Q$'s are categorized into the family of the $Q_2$-FFLO phases with an infinite rational vector $Q_2/(2n+1)$ ($n\!=\!1, 2, \cdots, \infty$). Thus, $Q$ and $\mathcal{M}$ in the ground state have a step structure, called the {\it devil's staircase} structure~\cite{chaikin,bak}. At $\mu _{\rm B}H \!=\! 0.552\Delta _{1,0}$, the $Q_2$-FFLO phase undergoes the first-order transition to the $Q_1$-FFLO phase with the shorter FFLO period.

%Although we here present the phase diagram in the case of $Q_2 > Q_1$, we also carry out the numerical calculation for the case of $Q_2 < Q_1$ ($|\Delta _1|>|\Delta _2|$) in the limit of high fields. The latter case inevitably has a critical field at which the $Q_2$ curve crosses with the $Q_1$ curve. The lower field regime is dominated by the minority ($\gamma=2$) band, leading to the emergence of the $Q_2$-FFLO phase, while the higher field in low $T$'s is occupied by the $Q_1$-FFLO phase~\cite{takahashi}. 

KFe$_2$As$_2$ is a strongly Pauli-limited superconductor with 
$\alpha$-, $\beta$-, $\zeta$- and $\epsilon$-bands~\cite{burger,zocco}.
A minor (major) gap forms on the $\alpha$-, $\beta$-, and $\zeta$- ($\epsilon$) bands 
with a relatively larger (smaller) Fermi velocity, which leads to $Q_{\alpha \beta \zeta} \!<\!
Q_{\epsilon}$ or $Q_2\!<\!Q_1$ in the present context. 
Note that the $Q_{2}<Q_1$ case inevitably has a critical field at which the $Q_2(H)$ curve crosses $Q_1(H)$. The lower-field regime with $Q_2 \neq 0$ and $Q_1=0$ is dominated by the minor band, in which the $Q_2$-FFLO phase is stabilized. The higher-field regime with $Q_2<Q_1$ turns to the $Q_1$-FFLO phase via the first order transition~\cite{takahashi}. Hence, the emergence of the $Q_2$-FFLO phase is a generic feature in two-band Pauli-limiting superconductors, but the $Q_2$-FFLO phase is not divided into subphases and the devil's staircase is absent, when $Q_2/Q_1\lesssim 1$. To fully understand KFe$_2$As$_2$, however, we have to take account of the orbital effect. The interplay between vortices and FFLO states in multiband systems remains as a future problem. Note that the generalized WHH approach has recently made the Pauli-limit effect of KFe$_2$As$_2$ questionable~\cite{kogan2012}.

%-------------------- Conclusions.
{\it Conclusions.---}
We have examined the multiband effects on FFLO phases and revealed that generic and nontrivial features are absent in a single-band case. Our calculation is based on a minimal model extended from a canonical 1D FFLO Hamiltonian~\cite{machida1984}. We have demonstrated that the FFLO phase diagram in the $H$ vs $T$ plane is divided into two main subphases by the first order transition, where the $Q_2$-FFLO phase in the lower-$H$ regime is further subdivided, giving rise to a devil's staircase in physical quantities. Yet, remarkably, the tricritical Lifshitz point $L$ at $T_{\rm L}/T_{\rm c} \!=\! 0.561\dots$, where the normal, FFLO, and uniform BCS phases meet~\cite{chaikin,hornreich}, is invariant even in the multiband case, as shown in Fig.~\ref{fig:phase}, independent of any of the parameters $g_{12}/g_{11}$, $g_{22}/g_{11}$, and $\mu _1/\mu _2$~\cite{takahashi}. This is a generic feature observed in various systems~\cite{machidaISDW,fujita,machidastripe,machida2005,TM2007}. The present findings on the FFLO state are also applicable to imbalanced superfluids with two chains \cite{sun1,sun2} in ultracold atoms.

\begin{acknowledgments}

This work was supported by JSPS (Grant Nos.~21340103, 23840034, and 25800199) and ``Topological Quantum Phenomena''  (No.~22103005) KAKENHI on innovation areas from MEXT.

\end{acknowledgments}
%This work was supported by JSPS
%(No.~2074023303, 2134010303 and 22540383) and the MEXT KAKENHI (No.~22103002 and
%No.~22103005).  
%``Topological Quantum Phenomena'' (No.~22103005) KAKENHI on innovation areas from MEXT.
% 2074023303: Mizushima (until Mar. 2012)
% 21340103: Machida 
% 25800199: Mizushima (Watate-B)

%---------- References 

\end{document}